# Exchange Bias in [Co$_2$MnGe/Au]$_n$ , [Co$_2$MnGe/Cr]$_n$ and [Co$_2$MnGe/Cu$_2$MnAl]$_n$ Multilayers

K. Westerholt, U. Geiersbach and A. Bergmann
Institut für Experimentalphysik IV, Ruhr-Universität  D44797 Bochum, Germany

**Abstract**
We report structural and magnetic properties of multilayers composed of thin layers of the half metallic ferromagnetic Heusler compound Co$_2$MnGe and layers of Au, Cr and the Heusler compound Cu$_2$MnAl. The hysteresis loops measured at low temperatures reveal the existence of an exchange bias field H$_{EB}$ in all of these multilayers. For the [Co$_2$MnGe/Au]$_n$ multilayer system H$_{EB}$ is largest reaching up to 1 kOe at a temperature of 2 K. We characterize the exchange bias phenomenon in detail and show that it originates from a spin glass type of magnetic order for a thin interlayer at the interfaces. We discuss the results in the light of different models proposed for the explanation of the exchange bias effect.

**Corresponding author**
Prof Dr. K. Westerholt
Institut für Experimentalphysik IV, Ruhr-Universität, D 44780 Bochum
Email: Kurt.Westerholt@ruhr-uni-bochum.de
Tel.: 0234/3223621

# Introduction

The growth and the magnetic properties of thin films of Heusler alloys with the general formula $A_2BX$ denoting four interpenetrating fcc sublattices occupied by A, B and X-atoms (A=Cu, Co, Ni…, B=Mn, Fe..., X=Al, Ge, Si…)[1], attracted considerable interest in recent years. These Heusler alloys provide several compounds which, following electronic energy band structure calculations, should be half metallic ferromagnets i.e. possess a 100% spin polarization at the Fermi level[2]-[4]. This property is very rare for intermetallics and until now among the huge number of intermetallic ferromagnetic compounds only several Heusler compounds are predicted to have this unique property. A complete spin polarization at the Fermi level is very attractive for the application in spin dependent electron transport as e.g. giant magneto resistance (GMR) or tunnelling magneto resistance (TMR) or for the injection of a spin polarized current into semiconductors[5].

The ferromagnetic Heusler half metals known from band structure calculations are the compounds PtMnSb and NiMnSb[4] (so called half Heusler compounds since one of the A-sublattices is empty) and the compounds $Co_2MnSi$, $Co_2MnGe$ and $Co_2MnSn_{1-x}Sb_x$[2],[3]. Single thin films and thin film heterostructures of PtMnSb and NiMnSb have been studied by several groups during the last decade[6],[7]. However, until now TMR- or GMR-elements using these Heusler compounds show only moderate performance[8],[9], thus doubts arose about the full spin polarization for these compounds. However, it must be kept in mind that the narrow gap in the minority spin band at the Fermi energy in the Heusler compound is induced by symmetry[2],[3] and deviations from the fully ordered $L2_1$-symmetry of the unit cell will partly fill up this gap with electronic states. Perfect site order in the Heusler unit cell is difficult to achieve even in bulk single crystals[1]. In thin film heterostructures, which must be processed at rather low temperatures in order to prevent excessive interdiffusion, site disorder can hardly be avoided completely. It is not clear until now, to what extend this site disorder will degrade the full spin polarization at the Fermi energy.

The half metallic ferromagnetic Heusler compounds $Co_2MnSi$, $Co_2MnGe$ and $Co_2MnSn_{1-x}Sb_x$ found little attention in the experimental literature until now. We recently have shown that homogeneous thin films of these materials with good structural quality and high magnetic moments can be grown[10]. Independent of us, other groups have also started experimental studies on thin films based of $Co_2MnSi$[11] and $Co_2MnGe$[12]. The magnetoresistance of spin valve devices composed of two layers of $Co_2MnGe$ with Cr or V-interlayers has been studied recently[13], only a small GMR-effect was observed.

Here we report on our investigations of multilayers combining very thin layers of the $Co_2MnGe$-phase with thin layers of Au-, Cr- and the second Heusler phase $Cu_2MnAl$. The main aim of this study is a characterization of the magnetic properties of $Co_2MnGe$ films in the limit of very small thicknesses of the order of the unit cell lattice parameter. In this thickness range the interfaces become increasingly important and one gets informations about the development of the magnetic moments, ferromagnetic Curie temperatures and the change of site disorder with decreasing thickness. Since in applications like the GMR and the TMR usually very thin magnetic films are needed[14],[15], this should give a more realistic basis for judging the potential of the Heusler alloys in spin dependent transport than the study of thick films.

Originally one of our motivations for studying the $Co_2MnGe$-based multilayers was to search for an oscillatory interlayer exchange coupling which exists in nearly all multilayers of the ferromagnetic transition metals with non ferromagnetic interlayers in the thickness range between 0.7 nm and 3 nm for the non-magnetic interlayer[16]. However, to this end our results were negative until now, we found no indications of an oscillatory exchange interaction. Instead, we observe that at low temperature the ferromagnetic hysteresis loops are shifted from the symmetric position on the magnetic field axis, a phenomenon called exchange bias and well known in thin film magnetism since 40 years[17]. An exchange bias field is normally as-sociated with ferromagnetic/antiferromagnetic (f/af) thin film systems as e.g the classical system (Co/CoO)[18] or (FeNi/FeMn)[19]. Mainly motivated by the fact that the exchange bias field has found a very important technical application in spin valves used for data read-out in computer storage technology, the exchange bias problem was re-discovered in recent years and new systematic theoretical and experimental studies enlightened the subject substantially (see ref [20],[21] for recent reviews). However, a quantitative understanding of the exchange bias phenomenon is still lacking and the theoretical models proposed are partly controversial. The main problem in calculating the exchange bias field $H_{eb}$ comes from the fact that it is determined by the structure and atomic distribution at the (f/af)-interface, which usually is not known in detail for real systems. We report here on the observation of an exchange bias effect in multilayers $[Co_2MnGe/Au]_n$, $[Co_2MnGe/Cr]_n$ and $[Co_2MnGe/Cu_2MnAl]_n$, where at the first glance the effect seems surprising since there are no antiferromagnetic layers present.

## Preparation and Experimental

The thin films used for the present study were deposited by dual source rf-sputtering with the substrate at a temperature of 300°C during the film growth. The base pressure of the sputtering system was $5 \cdot 10^{-8}$ mbar during the sputter deposition, as working gas we used pure Ar at a pressure of $5 \cdot 10^{-3}$ mbar. The sputtering rate was 0.04 nm/s for $Co_2MnGe$ and $Cu_2MnAl$, 0.06 nm/s for Au and 0.03 nm/s for Cr. A systematic change of the process parameters showed that these values gave the best structural results. The Heusler alloy films were sputtered using polycrystalline, stoichiometric, single phase Heusler alloys targets of 10 cm diameter, which we have prepared by high frequency melting of the components in high purity graphite crucibles. The commercial Au- and Cr-targets had a purity of 99.99 at.%.

All multilayers of the present study were grown on equipolished a-plane sapphire substrates, which were carefully cleaned and ion beam etched prior to deposition. During the sputtering process the substrates were moved automatically between the two targets of the dual source deposition chamber. After finishing 30 periods of the multilayers we deposited a 2 nm thick Au-cap layer at room temperature for protection against oxidation. We usually prepared series of 10 multilayers simultaneously within the same run with either the thickness of the $Co_2MnGe$-layer kept constant and the thickness of the other metal varied or vice versa. The thickness covered typically a range between 1 nm and 3 nm for each component.

The structural characterization of all samples was carried out by a thin film x-ray spectrometry using Cu-$K_\alpha$-radiation. The x-ray study combined small angle reflectivity, out-of-plane Bragg scans and out-of-plane rocking scans. For selected samples in-plane rocking scans at glancing incidence were taken by a rotating anode x-ray spectrometer. The magnetic measurements were performed by a commercial SQUID magnetometer (Quantum Design MPMS system).

## Results
### x-ray characterization

In Fig.1a we show the small angle x-ray reflectivity scan of a $[Co_2MnGe(3nm)/Au(3nm)]_{30}$ sample with a nominal thickness (as calculated from the sputtering rate) of 3 nm for both components. Above the critical angle for total reflection $\Theta_c$ the superstructure gives rise to Bragg peaks superimposed on the Fresnel-reflectivity. We observe sharp superlattice reflections up to 5$^{th}$ order, revealing good interface quality and low fluctuations of the layer thickness. From the reflectivity peak of order l at the angle $\Theta_l$ one can calculate the superlattice periodicity $\Lambda$ by using the relation[22]

$$\Lambda = \lambda / (2(\sqrt{\Theta_l^2 - \Theta_c^2}\,)) \qquad (1)$$

From a fit we get Λ=6.1 nm in good agreement with the nominal thickness. From simulations of the reflectivity curves using the Parratt formalism[23] we estimate an interface roughness of about 0.4 nm.

Next we performed out-of-plane Bragg scans over the total angular range. They revealed that all multilayers of the present study grown on $Al_2O_3$ a-plane possess a pure out-of plane texture which is (110) for the $Co_2MnGe$-, $Cu_2MnAl$- and Cr-layers, and (111) for the Au-layers. Fig.1b depicts the peak structure of the [$Co_2MnGe$(3nm)/Au(3nm)]$_{30}$ multilayer close to the fundamental Au-(111)- and $Co_2MnGe$-(220) Bragg peak. Besides the fundamental Bragg peak from the average lattice, the multilayer exhibits a rich satellite structure caused by the chemical modulation of the multilayer. Satellites up to the order l=+5 and l=–6 can be resolved, proving coherently grown superstructures in the growth direction with very sharp interfaces. From the position of the satellite peaks one can derive the superstucture periodicity from the separation Δ(2Θ) of the satellite of order l from the fundamental Bragg peak using [22]

$$\Lambda = \lambda / (l \cdot 2 \cdot \Delta\Theta \cdot \cos(\Theta)) \qquad (2)$$

From this relation we get a superlattice period of 5.8 nm, in good agreement with the value derived from the small angle x-ray reflectivity.

From the width of the satellite peaks at half maximum (FWHM) Δ(2Θ) we can derive the out-of-plane coherence length of the superstructure $D_{coh}$ using the Scherrer equation[22]

$$D_{coh} = \lambda / (\Delta(2\Theta) \cdot \cos(\Theta)) \qquad (3)$$

We estimate $D_{coh}$=70 nm i.e. comprising more than 10 superlattice periods. The fundamental Bragg peak in Fig.1b is positioned at 2Θ=39.5° i.e. closer to the Au (111) Bragg peak at 2Θ=38.5° than to the $Co_2MnGe$ (220) peak at 2Θ=44.2°. This shows that the $Co_2MnGe$-layers are elastically expanded out-of-plane and compressed in-plane thus taking a tetragonal distortion. For the sample in Fig.1 we also have performed in-plane Bragg scans in order to characterize the crystallinity in-plane. These scans revealed that the sample has a broad distribution of Bragg peaks and thus is polycrystalline in-plane. Thus the sample must be characterized as a strongly textured multilayer rather than a superlattice. This statement holds for all multilayers of the present study.

In Fig.2a,b we present the small angle reflectivity scan and the large angle Bragg scan of the [$Co_2MnGe$(3nm)/$Cu_2MnAl$(3nm)]$_{30}$ multilayer. In the reflectivity one finds sharp superstucture Bragg peaks up to the 4$^{th}$ order indicating a good quality of the layered structure with an interface roughness of about 0.5 nm. From a fit of the reflectivity curve we determine a superlattice periodicity Λ=6.2 nm, in good agreement with the nominal thickness. The Bragg scan close to the (220) peak exhibits one fundamental superlattice reflection at 2 Θ=43.2° and two weak satellite peaks at a distance Δ(2Θ) ≈1.6º giving a superlattice periodicity of 6.1 nm. From the FWHM of the satellite peaks we estimate an out-of-plane structural coherence length $D_{coh}$ of about 18 nm thus the superstructure in the growth direction is coherent over about 3 periods. The fundamental (220) Bragg peak of the multilayer is positioned at 2Θ=43.5º i.e. approximately in the middle of the (220) Bragg peak position of the $Co_2MnGe$ phase (2Θ=44.2º) and the $Cu_2MnAl$ phase (2Θ=42.6º), as expected when one assumes that the layers are in the relaxed state with both layers adopting their own lattice parameters.

In Fig.3 a,b we present the small angle reflectivity and the Bragg peak structure of the [$Co_2MnGe$(3nm)/Cr(3nm)]$_{30}$-multilayer. The superstructure Bragg peaks in the reflectivity curves can be resolved up to the 4$^{th}$ order and give a superstructure period of 5.8 nm. They are definitely less intense and sharp than for the multilayers in Fig 1 and 2, indicating an increased roughness of the interfaces. From a fit of the small angle reflectivity we estimate an interface roughness of about 0.7 nm. Although the $Co_2MnGe$-phase and bcc Cr possess a nearly perfect matching of the lattice parameter to within 0.2 %, the peak intensity of the fundamental Bragg reflection (220) (Fig.3b) is rather low and possesses no satellites. Thus the [$Co_2MnGe/Cr$]$_{30}$-multilayers are incoherent and polycrystalline, the structural quality is definitely worse than that of the other multilayer systems of the present study.

## **Magnetic properties**

In Fig.4 we have depicted the magnetization M(T) measured in an applied field of 1 kOe for multilayers [$Co_2MnGe(d)/Au(3nm)$]$_{30}$ for different thicknesses d of the $Co_2MnGe$ layer. With decreasing thickness of the $Co_2MnGe$ layers the saturation magnetization at low temperatures is continuously reduced compared to the bulk value $\sigma_s$=111 emu/g[1]. Simultaneously the ferromagnetic Curie temperature $T_c$ is lowered drastically from the bulk value of $T_c$= 829 K[1] to about $T_c \approx$320 K for d=2 nm and $T_c \approx$100 K for d=1.5 nm. For an even smaller thickness the $Co_2MnGe$-layers are no longer ferromagnetic.

In Fig.5 we have plotted the ferromagnetic saturation magnetization measured at 4 K versus the thickness of the $Co_2MnGe$ layers in the [$Co_2MnGe/Au$]$_{30}$ multilayers. One observes that the ferromagnetic saturation magnetization of the Heusler layers breaks down at a critical thickness d≈1.4 nm i.e. for a thickness of about two times the lattice parameter. In Fig.5 we have included the thickness dependence of the saturation magnetization of [$Co_2MnGe/V$]$_{30}$ and [$Co_2MnGe/Cr$]$_{30}$ multilayers. The [$Co_2MnGe/V$]$_{30}$ multilayers exhibit a thickness dependence similar to [$Co_2MnGe/Au$]$_{30}$ with a strong decrease of the magnetization close to d=1.5 nm. For the [$Co_2MnGe/Cr$]$_{30}$-system we have magnetization values for only two samples, but these samples follow the same trend with an even stronger reduction of the saturation magnetization. Thus the strong decrease of the saturation magnetization at low thickness seems to be a common feature of the Heusler films clamped between non-magnetic metallic layers which we also have observed for the $Co_2MnSn$-phase[24]. This can either indicate that the ordered L2$_1$-Heusler structure breaks down at typically d≈1.4 nm and below this thickness there is a random distribution of the Co, Mn and Ge-atoms at the A, B and X-sites and a concomitant vanishing of the ferromagnetism, or, alternatively, that at a length scale of about 0.7 nm there is an interdiffusion at the interfaces destroying the Heusler structure and the ferromagnetism.

In Fig.6 we present the main result of the present paper. We show examples of magnetic hysteresis loops measured for a [$Co_2MnGe(2.6nm)/Au(3nm)$]$_{30}$-multilayer after cooling in a magnetic field of H=+100 Oe. One sees that the magnetic hysteresis loop measured at 2 K is definitely asymmetric with a shift towards negative fields whereas the loop measured at 14 K is symmetric. This indicates that at low temperatures an unidirectional magnetic anisotropy field $H_{eb}$ (exchange bias field) exists which contributes a term

$$\Delta F = -\vec{H}_{eb} \cdot \vec{M} \qquad (4)$$

to the magnetic free energy F. We have observed a corresponding exchange bias field effect in virtually any [$Co_2MnGe/Au$]$_n$-multilayer we have prepared, independent of the thickness of the Au-layer. The exchange bias field originates from a spin coupling at the interfaces [20],[21], thus one expects that $H_{eb}$ should decrease with increasing thickness of the ferromagnetic layer d. In Fig.7 we have plotted the exchange bias field for [$Co_2MnGe/Au$]$_n$ multilayers and $Co_2MnGe/Au$ trilayers for different thickness of the Heusler layer d. Although with large error bars for the absolute values of $H_{eb}$, we clearly observe the expected linear dependence $H_{eb} \propto 1/d$.

In Fig.8a we have plotted the temperature dependence of $H_{eb}$. A non vanishing $H_{eb}$ only occurs below about 15 K, $H_{eb}$ is found to increase strongly with decreasing temperature. Since the exchange bias effect usually is expected in (f/af) thin layer systems and there are no antiferromagnetic layers present in our [$Co_2MnGe/Au$]$_{30}$-multilayers, the effect seems puzzling at the first glance. The key for the understanding of the exchange bias effect in [$Co_2MnGe$/Au]$_{30}$ -multilayers is the magnetic ordering of the thin interface layer between the $Co_2MnGe$ film and the Au-film. In Fig.8b we have plotted the temperature dependence of the magnetization of a [$Co_2MnGe$(1.2nm)/Au(3nm)]$_{30}$–multilayer. The thickness of the $Co_2MnGe$-layer d=1.2 nm for this sample is below the critical thickness for ferromagnetic order in Fig.5, thus the $Co_2MnGe$-layer is strongly disordered and/or mixed with Au from both sides and can be considered as one single interface layer. The M(T)-curve in Fig.8b reveals a definite difference between the field-cooled and the 0-field-cooled magnetization with a peak in the 0-field-cooled magnetization at a temperature of about 16 K. This onset of irreversible magnetic behaviour combined with low absolute values for the magnetization indicates spin glass order with a spin glass freezing temperature $T_f$ of about 16 K[25]. The spin glass freezing originates from competing ferromagnetic and antiferromagnetic interactions in the interface region, where we suppose that Co-, Mn, Ge- and Au-atoms form a disordered alloy. Comparing now the results shown of Fig.8a and Fig.8b one sees that the spin glass freezing temperature correlates well with the onset of the exchange bias field, strongly suggesting the spin glass freezing at the interfaces causes the exchange bias field for the ferromagnetic $Co_2MnGe$-layers.

A similar exchange bias phenomenon also exists in the other two multiplayer systems of the present study [$Co_2MnGe$/Cr]$_{30}$ and [$Co_2MnGe$/$Cu_2MnAl$]$_{30}$. In Fig. 9a we show the temperature dependence of $H_{eb}$ measured on a multilayer [$Co_2MnGe$(3nm)/Cr(3nm)]$_{30}$. In this multilayer system $H_{eb}$ is definitely smaller than in [$Co_2MnGe$/Au]$_{30}$ and at the same time exists up to much higher temperatures namely to up about 150 K. In Fig.9b we have plotted the field-cooled and 0-field-cooled magnetization of a sample [$Co_2MnGe$(1.5nm)/Cr(3nm)]$_{30}$ i.e. for a very small thickness of the Heusler layer and a low value for the saturation magnetization (see Fig. 5). We find indications of spin glass order with a freezing temperature of about 160 K, again coinciding with the onset temperature of the exchange bias effect in Fig.9a.

The temperature dependence of the exchange bias field of a multilayer from our [$Co_2MnGe$/$Cu_2MnAl$]$_{30}$-series is presented in Fig.10a. We find an exchange bias field of the same order of magnitude as observed in the [$Co_2MnGe$/Cr]$_{30}$-system but only up to temperature of 60 K. However, at variance to Fig.8a and Fig.9a there is a change of sign of the exchange bias field above 20 K and a small, negative $H_{eb}$ persists up to about 60 K. Actually the [$Co_2MnGe$/$Cu_2MnAl$]$_{30}$-multilayers are magnetically more complicated than the [$Co_2MnGe$/Au]$_{30}$- and the [$Co_2MnGe$/Cr]$_{30}$-multilayers, since the $Cu_2MnAl$-phase is magnetically ordered at low temperatures. $Cu_2MnAl$ in the $L2_1$-phase is ferromagnetic with a ferromagnetic Curie temperature of 650 K[1]. However, when prepared as a thin film at temperatures of 300 ºC, the $Cu_2MnAl$ phase grows essentially in the B2-stucture with complete site disorder between the Mn- and Al-positions. The B2-type phase of $Cu_2MnAl$ is known to exhibit low magnetic moments and spin glass order[26].

For a comparison with conventional (f/af)-exchange bias systems[21],[22], we next look at several characteristic features of the exchange bias phenomenon and for this purpose focus on the [$Co_2MnGe$/Au]$_{30}$-multilayer system which shows the largest effect. An important characterization of the exchange bias field, which allows conclusions about the sign of the exchange interactions responsible for $H_{eb}$, is the dependence on the amplitude of the cooling field[20]. Our result for a [$Co_2MnGe$/Au]$_{30}$-multilayer is shown in Fig.11. A strong increase of the $H_{eb}$ with decreasing cooling field indicates that the exchange interactions at the interface responsible for $H_{eb}$ are antiferromagnetic. The optimum field for obtaining the largest $H_{eb}$ is the field just sufficient to keep the $Co_2MnGe$-layers magnetically saturated. A field dependence of $H_{eb}$ is also known for classical (f/af)-exchange bias systems[20],[21], however the strong field dependence at low cooling fields observed for the [$Co_2MnGe$/Au]$_{30}$-multilayer system is unusual and indicates that many spins at the interface contributing to $H_{eb}$ are only weakly exchange coupled and can easily be rotated by an external magnetic field.

A further characteristic phenomenon of exchange bias systems is the instability of the exchange bias field, which depends on the time, the thermomagnetic history and the number of magnetization cycles (training effect)[20],[21]. This is caused by irreversible changes of the spin structures at the interfaces. Fig.12 shows how the hysteresis loops of [$Co_2MnGe/Au$]$_{30}$ multilayer changes after repeated cycling the magnetic field at a temperature of 2 K. We observe a relaxation of the exchange bias field with increasing number of cycles, after the first magnetization reversal $H_{eb}$ decreases by about 50%. Compared to (f/af) exchange bias systems the relaxation in our system appears to be unusually strong and in this respect resembles the (Co/CoO) system for the case of very thin CoO-layers[27].

Other examples of a relaxation of $H_{eb}$ are depicted in Fig. 13a,b. In Fig. 13a the sample has been field-cooled in a high magnetic field of +4T. One sees that for the first cycle the hysteresis loop is not closed and shifted in the direction of positive magnetization. This can be naturally attributed to the thermoremanent magnetization of the spin glass phase at the interface which is frozen in when cooling in a large positive magnetic field. Upon the magnetization reversal of the ferromagnetic film this irreversible magnetization is completely removed. A peculiar relaxation of $H_{EB}$, which to our knowledge has not been observed in (f/af)-exchange bias systems, is shown in Fig.13b. There we have depicted hysteresis loops obtained after field-cooling in a field of +40 kOe and then driving the magnetic field to -40 kOe. We observe an irreversible change of the magnetization towards negative values and, interestingly, a change of sign of $H_{eb}$. Directly after field-cooling from +40 kOe there is an exchange bias field $H_{eb}$ =+700 Oe, after driving the field to –40 kOe we have $H_{eb}$= -250 Oe. This irreversible negative magnetization is unstable and vanishes after the second field reversal with a recovery of the positive value of $H_{eb}$. The relaxation behaviour in Fig.13a,b clearly demonstrates that the thermoremanent magnetization of the spin glass phase and the exchange bias field are intimately coupled.

# Discussion

We now come to an interpretation of the exchange bias phenomenon observed for the $Co_2MnGe$-based multilayers. The characteristic experimental feature are very similar to those observed in standard (f/af)-exchange bias systems[20],[21]. However, in our multilayer system there are no antiferromagnetic layers present and we have concluded that very thin interface layers with spin glass type of magnetic order cause the exchange bias phenomenon.

Different microscopic theoretical models of the exchange bias phenomenon have been developed in recent years, but refer to (f/af)-exchange bias systems only. In ref.[28] it is assumed that domain walls in the antiferromagnet perpendicular to the (f/af)-interface are formed giving rise to random fields which sum up to the macroscopic exchange bias field $H_{eb}$ observed experimentally. The model proposed in ref. [29] is based on the existence of domain walls in the antiferromagnet parallel to the (f/af)-interface. The domain state model[18],[30], mainly developed from computer simulations of (f/af)-exchange bias systems, explains the exchange bias field by the net magnetization of small domains formed in the bulk of the antiferromagnetic film. Since all these models take the (f/af)-interface explicitly into consideration, they cannot be applied to our systems directly.

We think that the characteristic features of spin glass freezing[25] provides a straightforward explanation for the exchange bias phenomenon which we have observed. A spin glass can freeze with a thermoremanent magnetization (TRM) pointing in any direction. The similarity of the temperature dependence of the TRM of the spin glass and the exchange bias field suggests that the exchange bias field originates from the spin glass freezing process at the ferromagnet/spinglass (f/sg)-interface. The magnetic free energy of the spin glass phase at the interfaces will be minimized by optimising the spin directions with respect to the exchange coupling to the ferromagnetic layer, the exchange interactions within the spin glass and the external magnetic field. The exchange coupling to the ferromagnetic layer saturated in a cer-tain direction manifests itself as an unidirectional macroscopic anisotropy field, the exchange bias field $H_{eb}$. We think that the random field model for the exchange bias worked out in ref.[28] but modified by replacing the af-domain walls by the spin glass phase and identifying the random fields in the domain walls by the random fields at the (f/sg)-interface gives a reasonable basis for a theoretical description of the exchange bias effect in our multilayers. What seems rather unusual in the exchange bias phenomenon observed here compared to conventional (f/af)-exchange bias systems is the strong relaxation and the strong field dependence of $H_{eb}$. This probably reflects the broad distribution of activation energies in the spin glass with a tail extending to very low energies.

We have observed an exchange bias field in all $Co_2MnGe$-based multilayers which we have studied until now, with the exception of the system $[Co_2MnGe/V]_n$[24]. The qualitative features of the exchange bias phenomenon are similar for all these multilayers, however the non-magnetic layers determine quantitative details as e.g. the magnitude of $H_{EB}$ and its onset temperature, which differ strongly from system to system. This shows that the exchange interactions essential for the formation of the spin glass phase at the interfaces are introduced by the non-magnetic layers.

The $[Co_2MnGe/Au]_{30}$-multilayer system combines the lowest onset temperature and the highest absolute values of $H_{eb}$. This correlates with a superior layered structure which we have obtained for these multilayers with less interdiffusion, thus high structural coherence and sharp interfaces seem to favour the exchange bias effect. The multilayer system $[Co_2MnGe/Cu_2MnAl]_{30}$ is exceptional, since the spin glass order is not limited to the interfaces but exists for the whole $Cu_2MnAl$-film. In this respect the system resembles the very rare examples of (f/sg)-exchange bias systems reported in the literature[31]. Interestingly we have found that only for this system $H_{eb}$ changes sign at low temperatures. A change of sign in $H_{eb}$ seems difficult to explain in any of the theoretical models of exchange bias and we tentatively would attribute this effect to a spin reorientation at the interfaces at low temperatures.

# Summary and Conclusions

The main aim of the present paper was to shown that in $Co_2MnGe$-based multilayers an exchange bias effect exists which originates from the formation of a spin glass type of order in very thin interface layers. The phenomenon is quite common for multilayers with the $Co_2MnGe$-phase and the other fully spin polarized Co-based Heusler alloys[24]. We suggested that the common origin for the effect is intermixing and site disorder at the interfaces which introduces competing ferromagnetic and antiferromagnetic interactions and spin glass order.

Our findings are important in two respects. First, they indicate that the interfaces of the $Co_2MnGe$-layers between two non-magnetic layers are not ferromagnetic and thus will necessarily loose the full spin polarization of the ideal, perfectly ordered $Co_2MnGe$-phase. Since in spin dependent transport phenomena like GMR and TMR the interface magnetism is essential[14],[15], this will be a severe problem when trying to use the full spin polarization of the Heusler compounds for obtaining large GMR- or TMR-effects. We think that this might be an important reason for the rather disappointing performance of spin valve elements with fully spin polarized Heusler compounds which has been achieved until now[8],[9],[13]. One might suggest that lower preparation temperatures avoiding interdiffusion at the interfaces could help. However, this will inevitably lead to an increasing degree of site disorder and by this to a loss of the high spin polarization.

Second, we think that our observations of an exchange bias field $H_{eb}$ in the Co-Heusler based multilayers introduces new aspects in the actual discussion of the phenomenon of exchange bias in general. In the present theoretical models developed for (f/af)-exchange bias systems the spin structure at the interfaces is regarded as essential. The formation of a spin glass phase at the interface with typical spin glass dynamics, however, is usually not taken into consideration. Including roughness and chemical mixing at the (f/af)-interface, the formation of a spin glass type of magnetic order at the interfaces seems rather plausible to us. At variance to our multilayers, however, the spin glass interlayer in (f/af)-systems will be exchange coupled to the ferromagnetic layer on one side and to the antiferromagnetic layer on the other side thus the magnetic behaviour will be more complex. In this sense is seems tempting to interpret the exchange bias phenomenon observed in our $Co_2MnGe$-based multilayers as a precursor of the exchange bias phenomenon in conventional (f/af)-systems.

*Acknowledgements*
*The authors thank the DFG for financial support of this work within the SFB 491,*
*P. Stauche for the preparation of the alloy targets and S. Erd-Böhm for running the sputtering equipment.*

## References

(1) P.J.Ziebeck, K.R.A.Webster; Landolt-Börnstein New Series **III/19c** (1986)
(2) S. Ishida, T.Masaki, S.Fujii, S.Asano; Physica B **145** (1998) 1
(3) S. Ishida, T.Masaki, S.Fujii, S.Asano; J. Phys. Soc. Jpn. **64** (1995) 2152
(4) R.A. de Groot, F.M.Muller, P.G. van Engen, K.H.J. Bushow; Phys. Rev. Lett. **50** (1983) 2024
(5) J.H. Park, E.Vescovo, E.Kim, C.Kwon, R.Ramesh, T.Venkatesan; Nature **392** (1998) 794
(6) M. C. Kautzky, F. B. Mancoff, J.-F. Bobo, R. P. Johnson, R . L. White, and B. M. Clemens;
    J. Appl. Phys. **81** (1997) 4026
(7) R.Kabani, M.Terada, A.Roshko, J.S. Moodera; J. Appl. Phys. **67** (1990) 4898
(8) J.A.Caballero, A.C.Reilly, Y. Hao, J. Bass, W.P.Pratt Jr., F.Petroff, J.R.Childress;
    J. Mag. Mag. Mat. **198-199** (1999) 55
(9) C. Tanaka, J. Novak, J. S. Moodera; J. Appl. Phys, **81** (1997) 5515
(10) U. Geiersbach, A. Bergmann and K. Westerholt; J. Magn. Magn. Mat. **240**/1-3 (2002) 546
(11) M. P. Raphael, B. Ravel, M.A.Willard, S.F.Cheng, D.N.Das, R.M.Stout, K.M. Bussmann,
    J.H. Claassen, and V.G. Harris; Appl. Phys. Lett. **79** (2001) 4396
(12) T. Ambrose, J.J. Krebs, and G. A. Prinz; Appl. Phys. Lett. **76** (2000) 3280
(13) T. Ambrose, J.J. Krebs, and G.A. Prinz; J. Appl. Phys. **89** (2001) 7522
(14) For a review see e.g. B. Dieny; J. Magn. Magn. Mat. **136** (1994) 335
(15) J. S. Moodera, L. R. Kinder, T. M. Wong, and R. Mersevey; Phys. Rev. Lett. **74** (1995) 3273
(16) S.S.P. Parkin in "Ultrathin Magnetic Structures II" p. 148,
    B. Heinrich, J. A. C. Bland (editors), Springer Verlag, Heidelberg, 1994
(17) W. Meiklejohn and C.P. Bean; Phys. Rev. **102** (1956) 904
(18) P. Miltenyi, M. Gierlings, J. Keller, B. Beschoten, G. Güntherodt, U. Novak and K. D. Usadel;
    Phys. Rev. Lett. **84** (2000) 4224
(19) D. Mauri, E. Kay, D. Scholl, and J. K. Howard; J. Appl. Phys. **62** (1987) 2929
(20) J. Nogues and I, K, Schuller J. Magn. Magn. Mat. **192** (1999) 203
(21) A. E. Berkowitz and K. Takano; J. Magn. Magn. Mat. **200** (199) 552
(22) H. Zabel; Advances in Sol. St. Phys. **30** (1990) 197
(23) L. G. Parratt; Phys. Rev. **95** (1954) 359
(24) U. Geiersbach, A. Bergmann and K. Westerholt (in preparation )
(25) K. Binder and A. P. Young; Rev. of Mod. Phys. **58** (1986) 801
(26) R. C. Taylor and C. C. Tsuei; Sol. St. Comm. **41** (1982) 503
(27) M. Gruyters and D. Riegel; Phys. Rev. B **63** (2000) 52401
(28) A. P. Malozemoff; Phys. Rev. B **35** (1987) 3679
(29) D Mauri, H. C. Siegmann, P. S. Bagus, and E. Kay; J. Appl. Phys. **62** (1987) $_{30}$47
(30) U. Novak, A. Mistra, and K. D. Usadel; J. Appl. Phys. **89** (2001) 7269
(31) B. Aqktas, Y. Öner and H. Z. Durusoy J. Magn. Magn. Mat. **119** (1993) 339

# Figure captions

**Fig.1a,b**

(a) Small angle x-ray reflectivity of the multilayer [$Co_2MnGe$(3nm)/Au(3nm)]$_{30}$ and
(b) x-ray intensity pattern close to the (220)/(111) fundamental Bragg peak, The numbers in the figure denote the order of the satellites

**Fig.2a,b**

(a) Small angle x-ray reflectivity of the multilayer [$Co_2MnGe$(3nm)/$Cu_2MnAl$(3nm)]$_{30}$ and
(b) x-ray intensity pattern close to the (220) fundamental Bragg peak

**Fig.3a,b**

(a) Small angle x-ray reflectivity of the multilayer [$Co_2MnGe$(3nm)/Cr(3nm)]$_{30}$ and
(b) and x-ray intensity pattern close to the (220) fundamental Bragg peak

**Fig.4**

Magnetization versus temperature for multilayers [$Co_2MnGe$(d)/Au(3nm)]$_{30}$ with the thickness of the $Co_2MnGe$-layer d given in the figure

**Fig.5**

Relative saturation magnetization versus the thickness of the $Co_2MnGe$-layer for multilayers

[Co$_2$MnGe(d)/Au(3nm)]$_{30}$ (squares), [Co$_2$MnGe(d)/Cr(3nm)]$_{30}$ (triangles) and [Co$_2$MnGe(d)/V(3nm)]$_{30}$ (circles)

**Fig.6**
Magnetic hysteresis loops measured at 4 K and 14 K for the sample [Co$_2$MnGe(2.6nm)/Au(3nm)]$_{30}$ after field-cooling in H= 200 Oe

**Fig.7**
Exchange bias field versus the thickness of the Co$_2$MnGe-layer in [Co$_2$MnGe(d)/Au(3nm)]$_{30}$ multilayers. The cooling field was 200 Oe, the temperature of the measurement was 2 K

**Fig.8a,b**
**(a)** Exchange bias field versus temperature for the multilayer [Co$_2$MnGe(2.6nm)/Au(3nm)]$_{30}$ for a cooling field H=200 Oe.
**(b)** field-cooled and 0-field-cooled magnetization for the sample [Co$_2$MnGe(1.2nm)/Au(3nm)]$_{30}$ for a magnetic field of 500 Oe

**Fig.9a,b**
(a) Exchange bias field versus temperature for the multilayer [$Co_2MnGe$(3nm)/Cr(3nm)]$_{30}$ after field cooling in H=200 Oe and

(b) field-cooled and 0-field-cooled magnetization for the sample [$Co_2MnGe$(1.5nm)/Cr(3nm)]$_{30}$ for a magnetic field of 200 Oe

**Fig.10a,b**
(a) Exchange bias field versus temperature for a [$Co_2MnGe$(3nm)/$Cu_2MnAl$(3nm)]$_{30}$ multilayer after field cooling in H=200 Oe

(b) field-cooled and 0-field-cooled magnetization for the same sample in a magnetic field of 200 Oe

**Fig.11**
Exchange bias field versus the cooling-field for the sample [$Co_2MnGe$(2.6nm)/Au(3nm)]$_{30}$ measured at a temperature of 2 K

**Fig.12**
Magnetic hysteresis loops of the multilayer [$Co_2MnGe$(2.6nm)/Au(3nm)]$_{30}$ measured at 2 K after field cooling in H=2 kOe. The labels in the figure denote the number of the field cycle

**Fig.13a,b**
(a) Magnetic hysteresis loops of the multilayer [$Co_2MnGe$(2.6nm)/Au(3nm)]$_{30}$ measured at 2 K after field cooling in H=40 kOe. The labels in the figure denote the number of the field cycles.

(b) The same as in (a), only the field has been driven to –40 kOe during the first field cycle

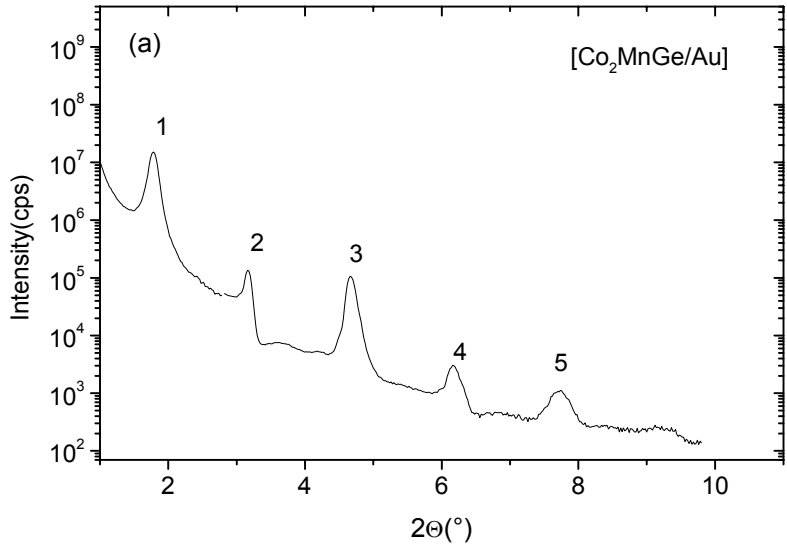

**Fig.1a**

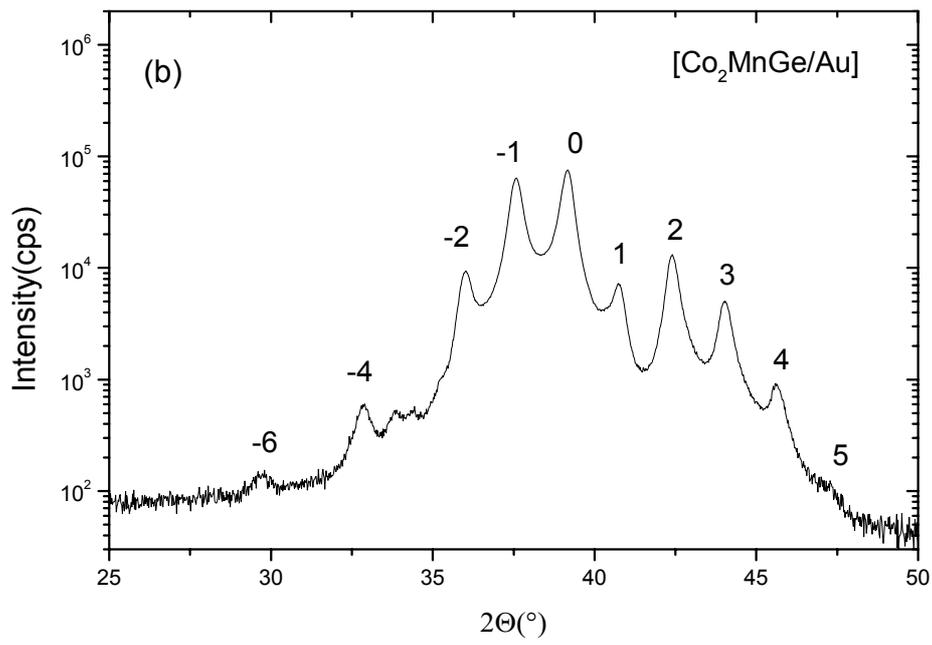

**Fig.1b**

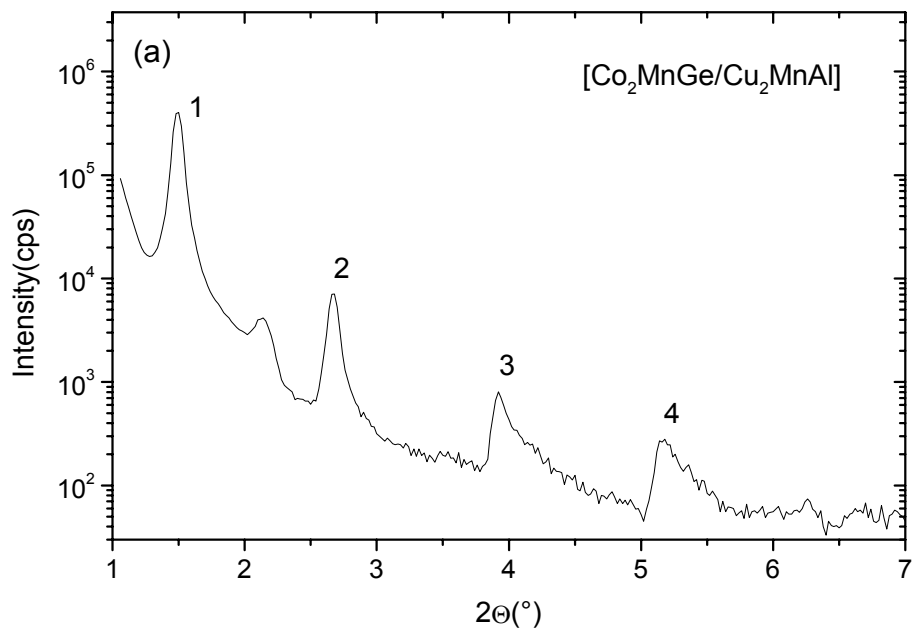

**Fig.2a**

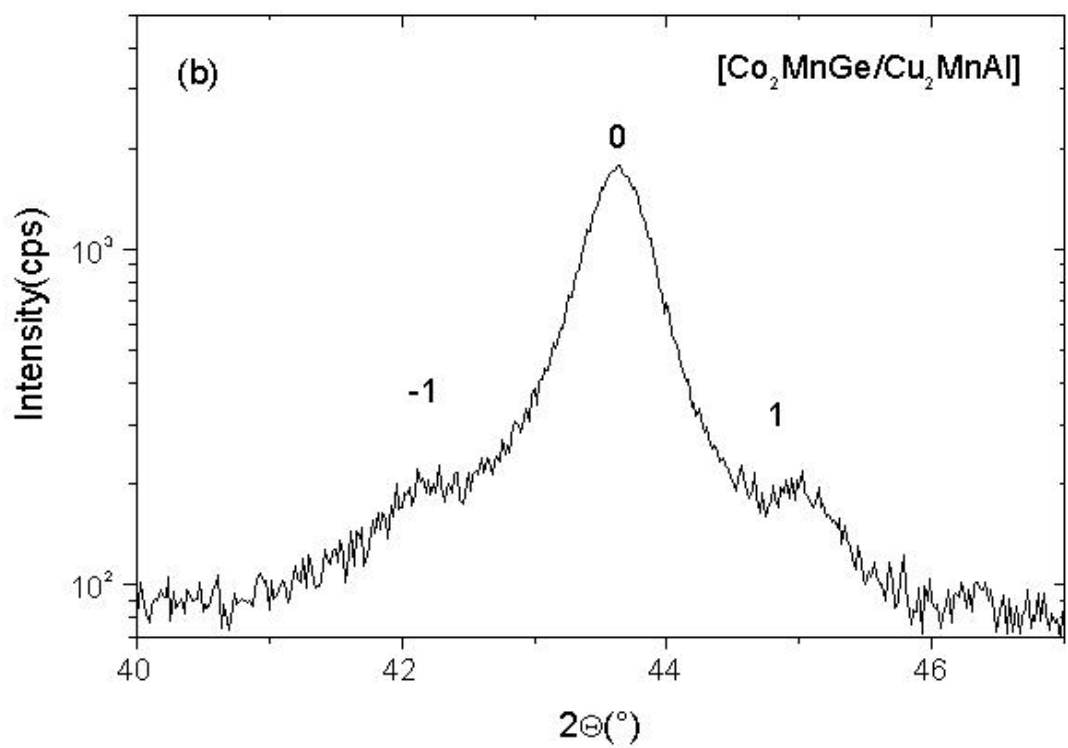

**Fig.2 b**

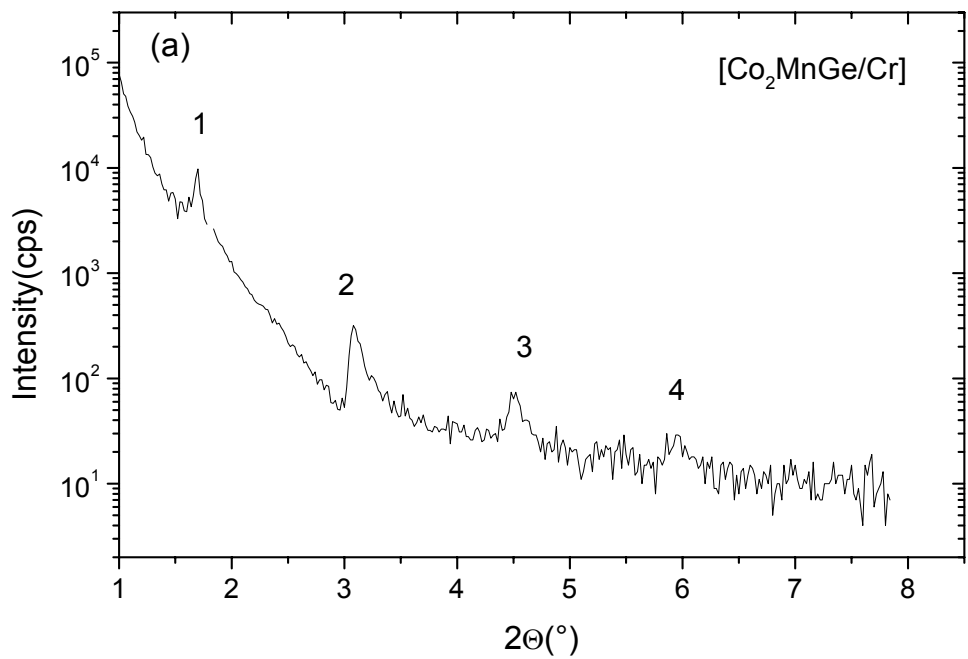

**Fig.3a**

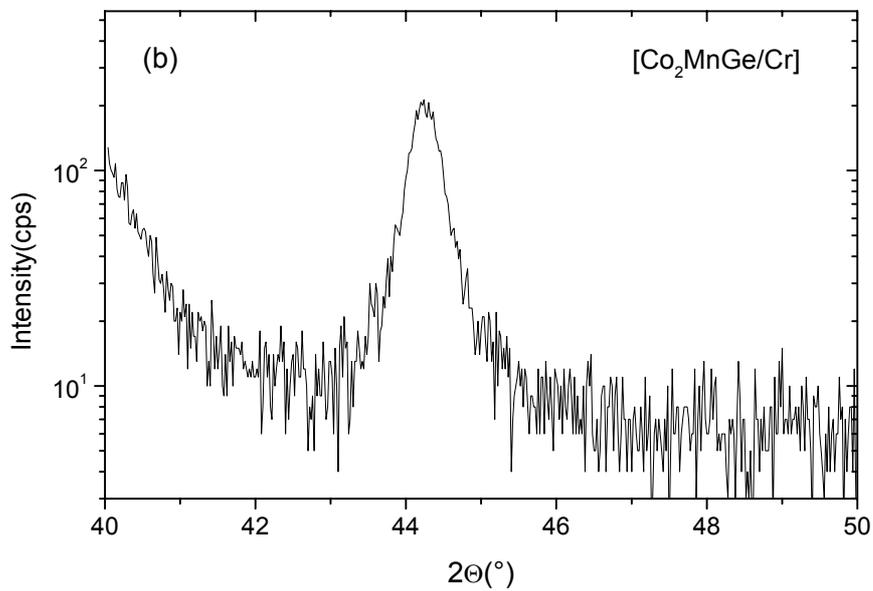

**Fig. 3b**

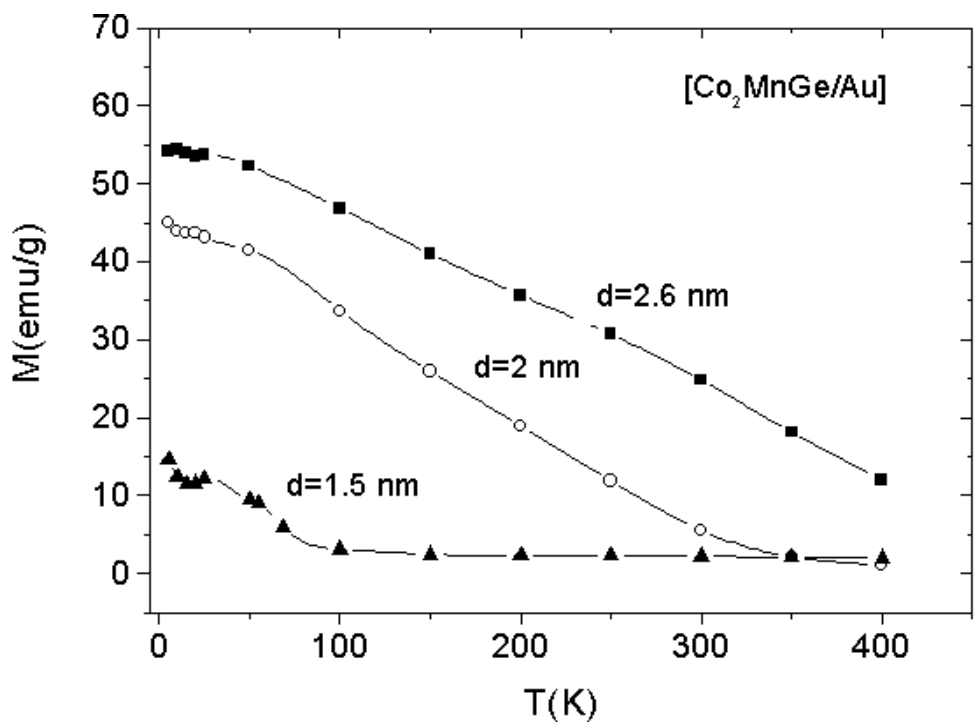

**Fig.4**

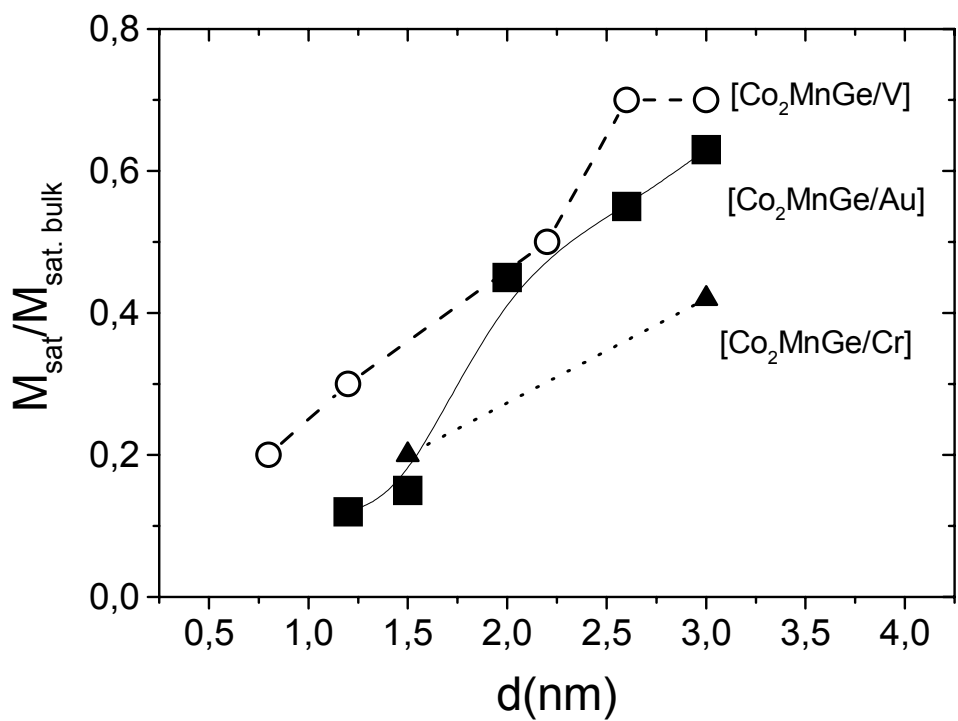

**Fig..5**

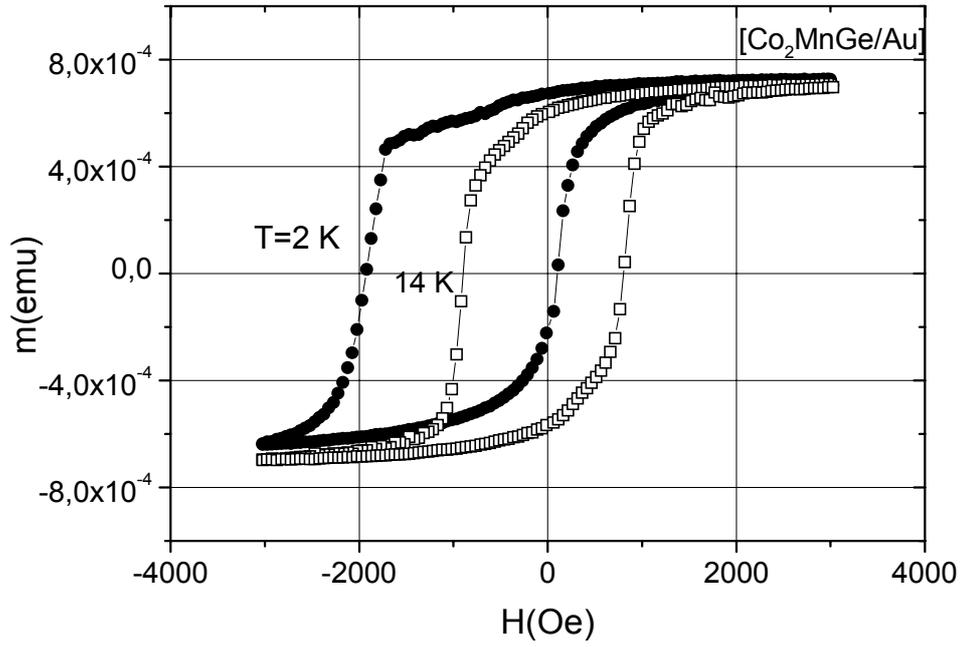

**Fig.6**

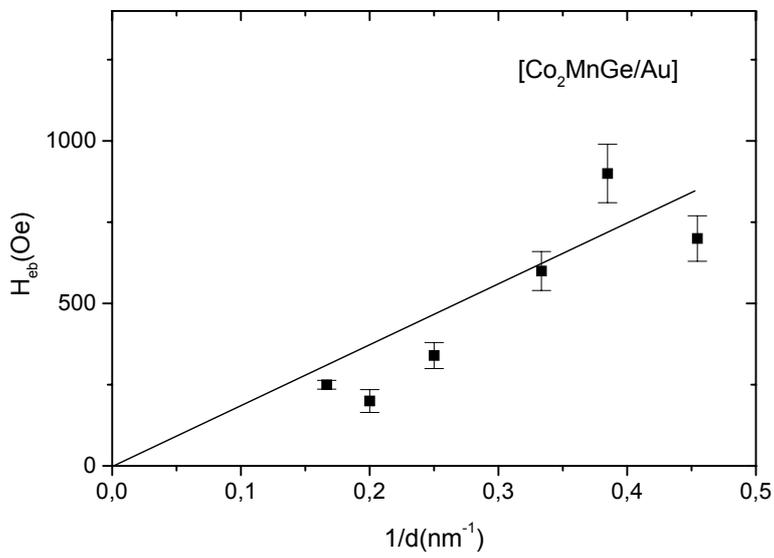

**Fig.7**

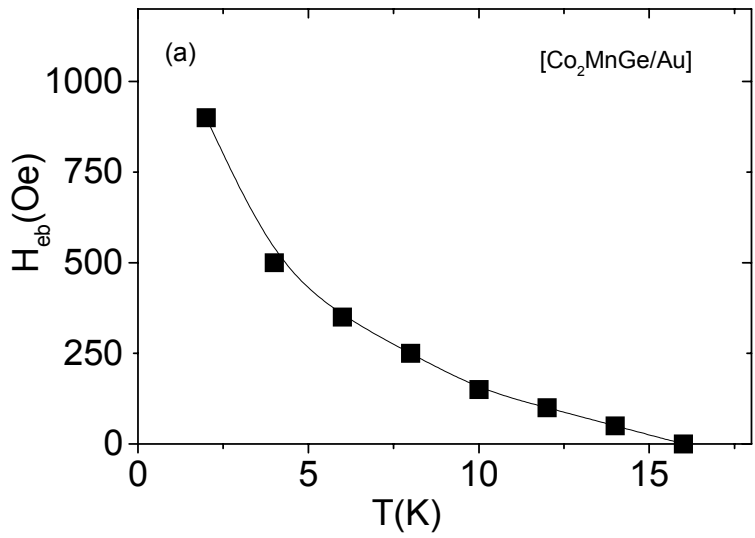

**Fig.8a**

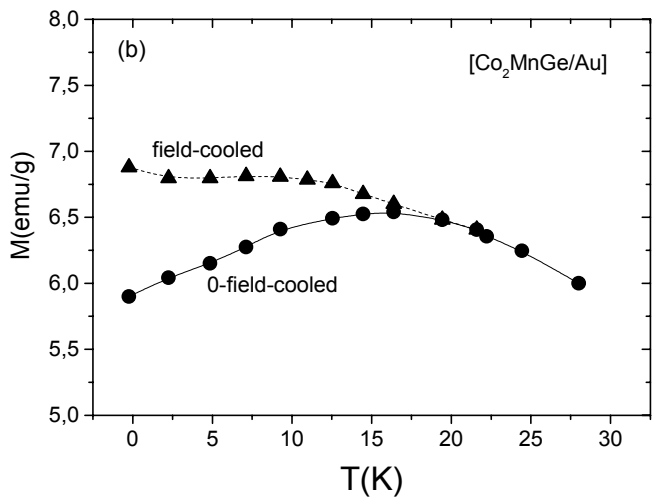

**Fig 8b**

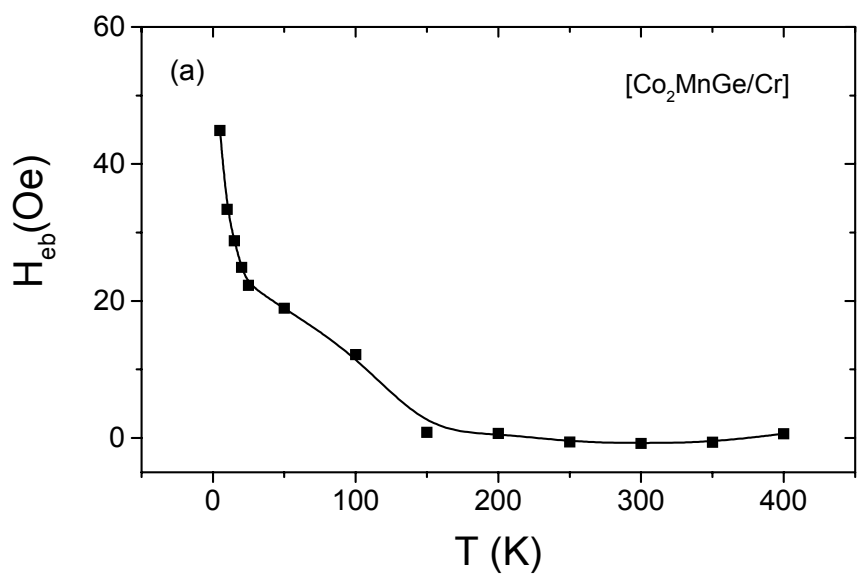

**Fig.9a**

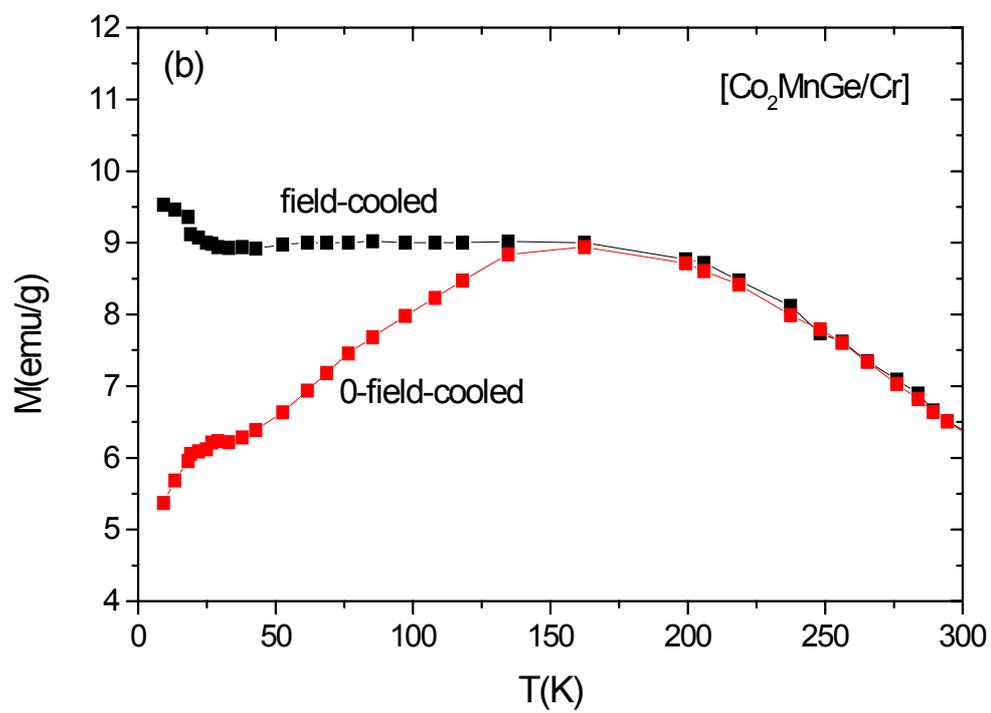

**Fig.9b**

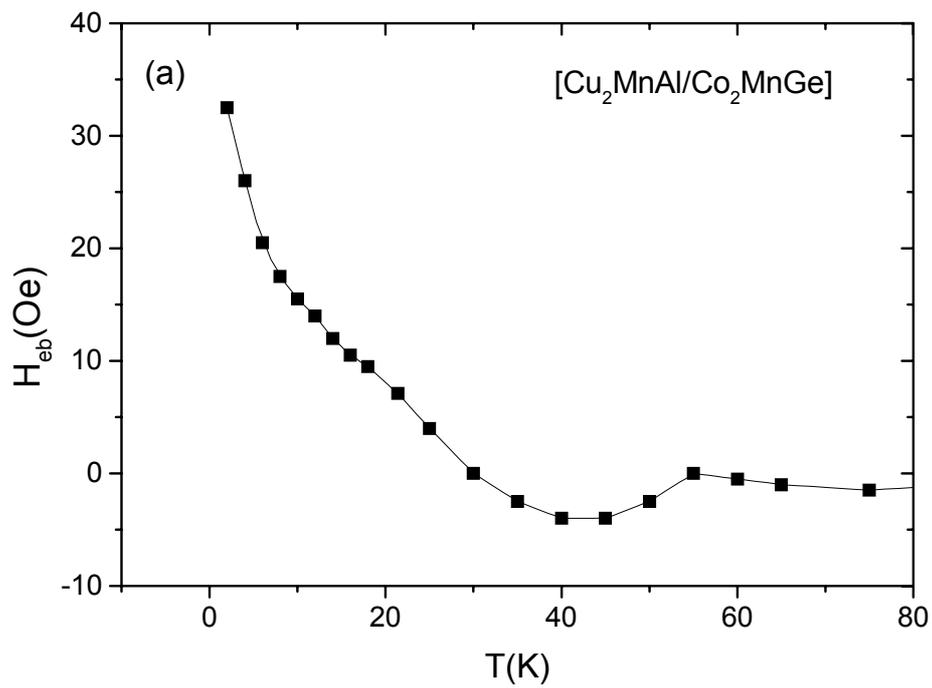

**Fig.10a**

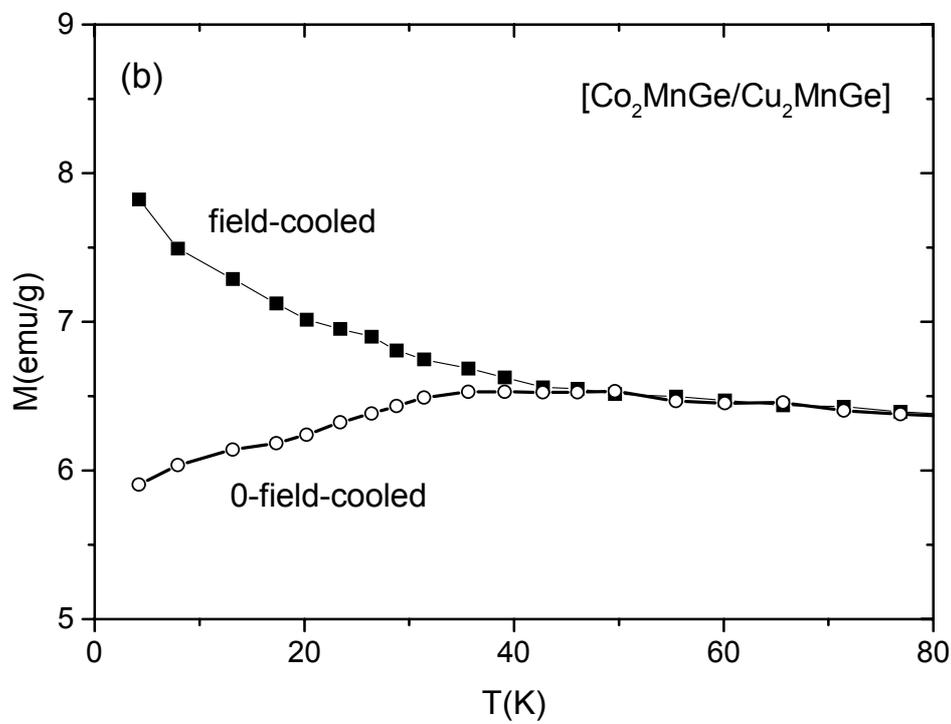

**Fig.10b**

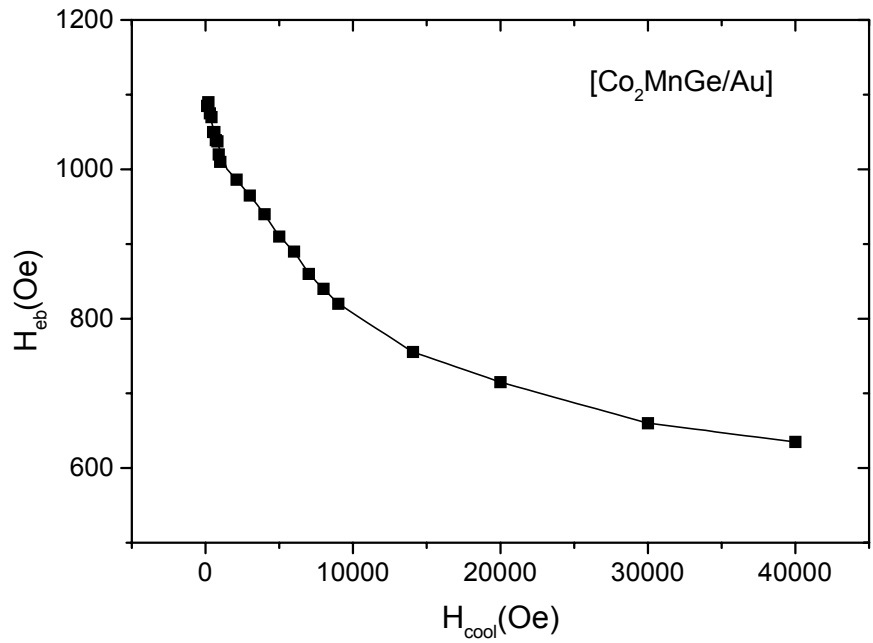

**Fig.11**

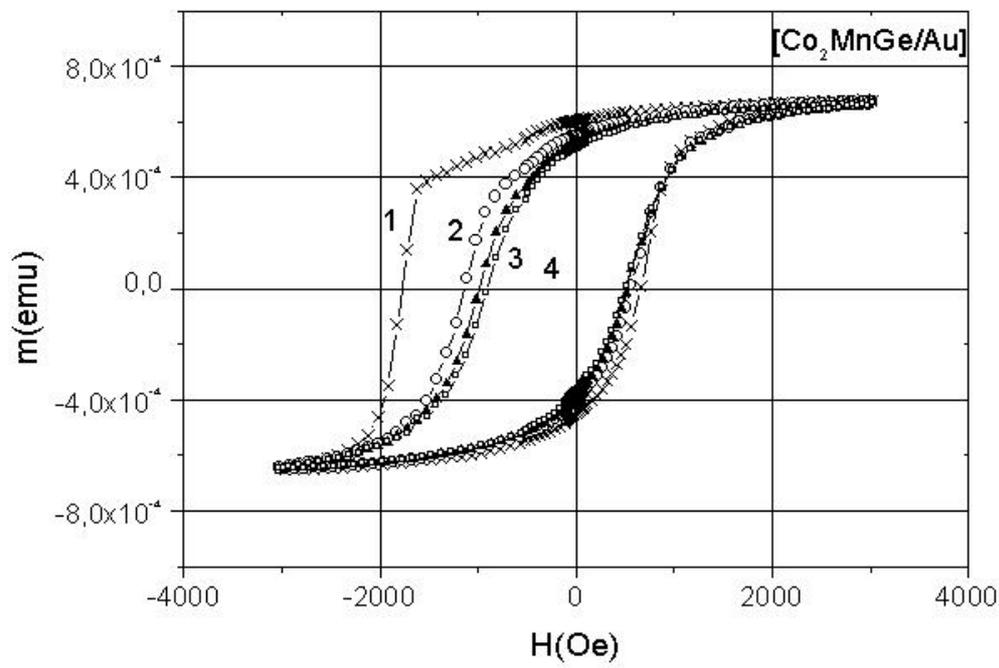

**Fig.12**

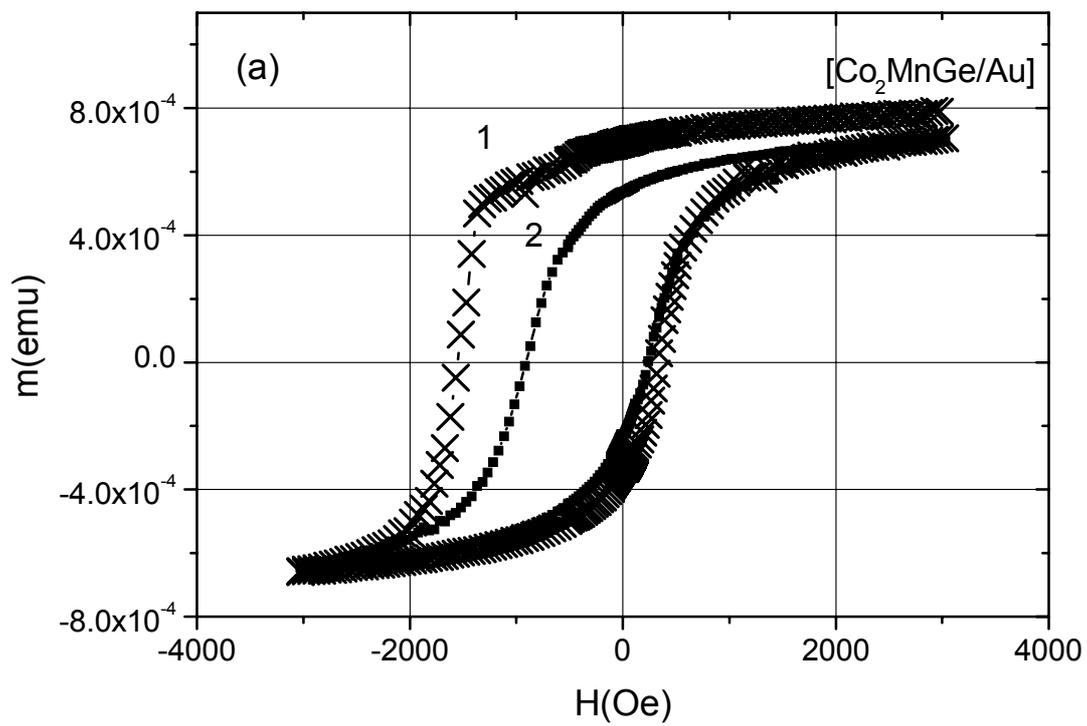

**Fig.13 a**

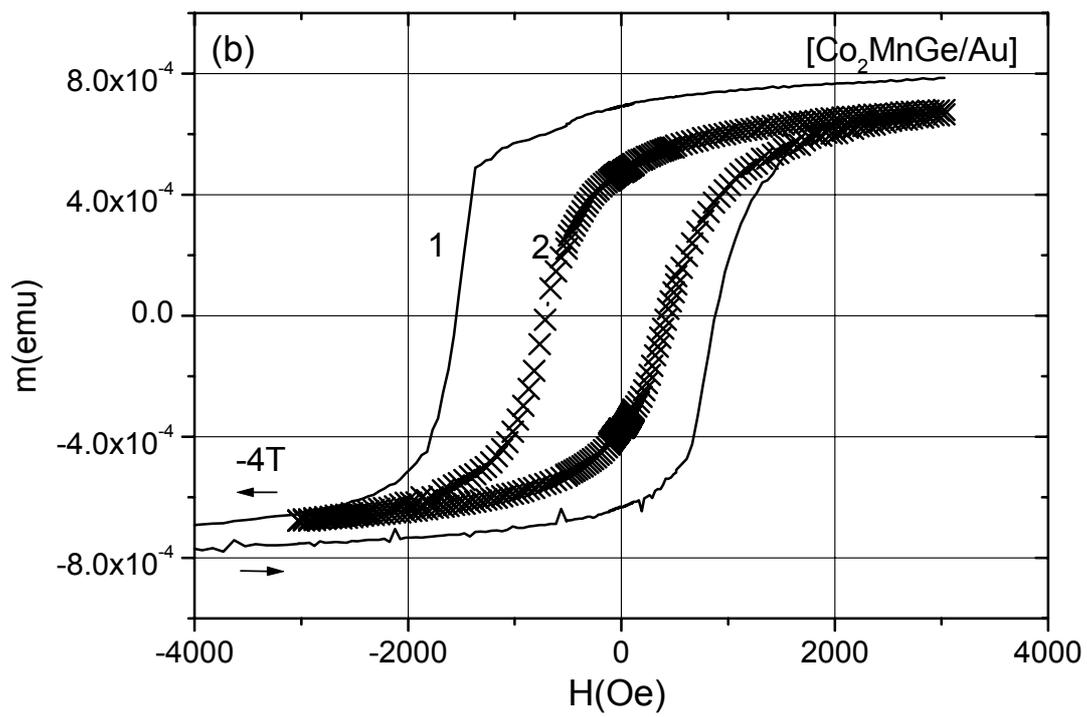

**Fig.13b**